\documentclass[pdftex,nofootinbib,superscriptaddress,aps,prfluids,notitlepage,longbibliography]{revtex4-2}
\usepackage{multirow,enumerate,epsfig,graphics,amssymb,amsmath,subeqnarray,mathrsfs,color,xcolor,epstopdf}
\usepackage{natbib}
\usepackage{color,epsfig,graphics,amssymb,mathcmd,amsmath,subeqnarray,graphicx,amsthm,mathrsfs} 
\usepackage[colorlinks=true, citecolor=red, linkcolor=blue ]{hyperref}
\usepackage{mathtools,bm} 
\usepackage{rotating,mathrsfs}
\usepackage{booktabs,subcaption,amsfonts,dcolumn}
\usepackage{multirow}
\usepackage[normalem]{ulem}
\usepackage{xcolor}

\makeatletter
\def\sgn{\mathop{\operator@font sgn}}
\def\threevdots{\vbox{\baselineskip1\p@ \lineskiplimit\z@
  \kern6\p@\hbox{.}\hbox{.}\hbox{.}}}
\makeatother

\begin{document} 
\title{A machine learning approach to the prediction of heat-transfer coefficients in micro-channels } 
\author{T. Traverso}
\email{ttraverso@turing.ac.uk}
\affiliation{The Alan Turing Institute, British Library, 96 Euston Road, London NW1 2DB, UK}
\affiliation{Department of Aeronautics, Imperial College London, South Kensington Campus, London SW72AZ, UK}
\author{F. Coletti}
\affiliation{Department of Chemical Engineering, Brunel University London, Kingston Lane, Uxbridge, Middlesex UB8 3PH, UK}
\affiliation{Hexxcell Ltd., Foundry Building, 77 Fulham Palace Rd., London, W6 8AF, UK}
\author{L. Magri}
\email{l.magri@imperial.ac.uk}
\affiliation{Department of Aeronautics, Imperial College London, South Kensington Campus, London SW72AZ, UK}
\affiliation{The Alan Turing Institute, British Library, 96 Euston Road, London NW1 2DB, UK}
\author{T. G. Karayiannis}
\affiliation{Department of Mechanical and Aerospace Engineering, Brunel University London, Kingston Lane, Uxbridge, Middlesex UB8 3PH, UK}

\author{O. K. Matar}
\affiliation{Department of Chemical Engineering, Imperial College London, South Kensington Campus, London SW72AZ, UK}
\affiliation{The Alan Turing Institute, British Library, 96 Euston Road, London NW1 2DB, UK}
\begin{abstract}
 The accurate prediction of the two-phase heat transfer coefficient (HTC) as a function of working fluids, channel geometries and process conditions is key to the optimal design and operation of compact heat exchangers. 
   Advances in artificial intelligence research have recently boosted the application of machine learning (ML) algorithms to obtain data-driven surrogate models for the HTC. For most supervised learning algorithms, the task is that of a nonlinear regression problem. Despite the fact that these models have been proven capable of outperforming traditional empirical correlations, they have key limitations such as overfitting the data, the lack of uncertainty estimation, and  interpretability of the results.
   To address these limitations, in this paper, we use a multi-output Gaussian process regression (GPR) to estimate the HTC in microchannels as a function of the mass flow rate, heat flux, system pressure and channel diameter and length. The model is trained using the Brunel Two-Phase Flow database of high-fidelity experimental data. 
   The advantages of GPR are  data efficiency, the small number of hyperparameters to be trained (typically of the same order of the number of input dimensions), and the automatic trade-off between data fit and model complexity guaranteed by the maximization of the marginal likelihood (Bayesian approach). Our paper proposes research directions to improve the performance of the GPR-based model in extrapolation. 
\end{abstract}

\maketitle 

\section{Introduction}

   The dissipation of high heat fluxes from small surface areas is one of the main challenges in the thermal design of electronic devices. Dissipation can be  achieved effectively by using two-phase flow boiling in small-to-micro diameter channels, a solution that has the advantage of dissipating high heat fluxes at nearly uniform substrate surface temperatures \cite{karayiannis2017flow}. The accurate prediction of the heat transfer coefficient (HTC) is key to the optimal design and operation of such compact heat exchangers. However, 
   due to the nonlinear behaviour in phase-change systems such as flow boiling, the accurate prediction of the HTC is challenging \cite{Gherhardt_2007}. To this end, physics-based models and empirical correlations have been developed as design tools for a large number of fluid flows and heat transfer configurations. However, the developed micro-scale correlations may not be extrapolated with confidence outside their applicability ranges, while few physics-based models have been developed. This is the case especially at the micro- and nano-scales, where the underlying physical phenomena are complex and remain relatively poorly understood  \cite{mahmoud2013heat}. 

A more flexible approach is to use data-driven numerical methods to perform regressions on the available data, and infer the HTC under different working conditions or for new heat exchanger designs. Recent advances in the field of artificial intelligence are making this option suitable for solving heat transfer problems \cite{Hughes2021,loyola2022machine}. Specifically, machine learning algorithms have the potential for extracting  trends from intricate datasets, generating robust optimization frameworks, and providing accurate prediction, often outperforming traditional methods \cite{hughes2021universal,zhou2020machine}. 
Despite their remarkable performance, commonly employed supervised ML models, such as support vector regression, random forest, and artificial
neural networks (ANN), present significant limitations. In fact, these models do not provide uncertainty quantification, may overfit data due to the large number of parameters, and might require tedious fine tuning of their hyperparameters \cite{Hughes2021}. Their predictions are  not readily interpretable, providing limited insight into understanding the data.

In this work, we train Gaussian process regression models to infer the two-phase HTC in micro-tubes as a function of five design parameters, and two working fluids, using part of the \emph{Brunel Two-Phase Flow database}. We demonstrate how this family of ML models present key features, which make them a promising method to perform this task. First, GPR are non-parametric models that are designed not to overfit the data, dramatically reducing the degree of arbitrariness introduced at the model design stage (here, reduced to choosing the appropriate kernel function). Moreover, the kernel parameters are often easily interpretable, providing a powerful tool for understanding the trends in the data, as discussed in section \ref{sec:2}. 
Second, GPR models are data-efficient becasue they require relatively small datasets  to achieve a good predictive performance. Finally, GPR models are based on a Bayesian framework, which naturally provides uncertainty estimation. Conversely, if the data uncertainty is known, the uncertainty information can be consistently incorporated into the model prediction.  
We end the paper by providing directions to enable the extrapolation to unseen scenarios.

\section{Gaussian Process Regression} \label{sec:2}
Gaussian process models, or simply Gaussian processes (GPs), are popular because they can be used in a regression framework to approximate  nonlinear functions with probabilistic estimates of their uncertainties \cite{swiler2020survey}. This section provides an overview of  Gaussian process regression (GPR). 

A Gaussian process is a collection of random variables $ \{ f(\textbf{x}) | \textbf{x} \in \mathcal{X} \subset \mathbb{R}^n \}$ for which, given any finite set of $N$ inputs $\textbf{X}=\{\textbf{x}_i \in \mathcal{X} | i=1, \dots, N \}$, the collection $f(\textbf{X}) \equiv \{ f(\textbf{x}_i) | i=1, \dots, N \}$ has a joint multivariate Gaussian distribution. This joint distribution is denoted as
\begin{equation}
    f(\textbf{X}) \sim  GP(m(\textbf{X}),  k(\textbf{X}, \textbf{X})  )  \label{prior}
\end{equation}
where $m(\textbf{X})$ is the $N$-dimensional vector with elements $m(\textbf{x}_i)= \mathbb{E}\left[ f(\textbf{x}_i) \right]$, i.e., the mean of $f$ at $\textbf{x}_i$, and $k(\textbf{X}, \textbf{X})$ is the $N$-by-$N$ covariance matrix defined as
\begin{equation}
      k_{ij} = k(\textbf{x}_i,\textbf{x}_j) = \mathbb{E}\left[ \left( f(\textbf{x}_i) - m(\textbf{x}_i) \right) \left( f(\textbf{x}_j) - m(\textbf{x}_j) \right) \right] \label{kij}
\end{equation}
The GP is  defined by the knowledge of its mean and covariance. 

Before regression is performed (i.e., before data is considered), the functions $m$ and $k$ reflect prior knowledge about the unknown function $f$. In this work, the form of the kernel $k(\textbf{x},\textbf{x}')$ is chosen a priori and then tuned by adjusting a finite set of continuously varying hyperparameters $\boldsymbol{\theta}$. This form of training is known as \textit{type II maximum likelihood approximation} or ML-II. 
The prior mean function is  assumed to be zero unless otherwise stated, that is $m(\textbf{x})=0$ for all $\textbf{x}$. 

For a GPR model, we consider a Gaussian process $f$ and a data set $D$. The data set comprises of a vector of $N$ noisy observations $\textbf{y}=[y_1, \dots, y_N]$ at the input locations $\textbf{X}=[\textbf{x}_1,\dots,\textbf{x}_N]$. These observations are obtained
from the true value $f(\textbf{x}_i)$ plus additive independent and identically distributed Gaussian noise $\epsilon$ with variance $\sigma_n^2$, that is
\begin{align}
    y_i &=  f(\textbf{x}_i) + \epsilon \\
    \epsilon &\sim \mathcal{N}(0,\sigma_n^2) 
\end{align}

The model is \emph{trained} by maximizing the marginal likelihood of the models hyperparameters ($\hat{\boldsymbol{\theta}}$) given the data, $p(\textbf{y} | \textbf{X}, \hat{\boldsymbol{\theta}})$. 
The models hyperparameters define the kernel, and the noise variance, i.e.,  $\hat{\boldsymbol{\theta}} = \{\boldsymbol{\theta} , \sigma_n^2 \}$. The resulting optimization problem is  
\begin{equation}
   \arg \max_{\hat{\boldsymbol{\theta}}} p(\textbf{y} | \textbf{X}, \hat{\boldsymbol{\theta}})  
 \label{training} 
\end{equation}
which is solved by using a gradient based approach in the GP framework GPy \cite{gpy2014}.
Finally, by conditioning the joint distribution on the training data and the testing inputs $\textbf{x}_*$, we derive the predictive distribution
\begin{equation}
    \textbf{f}_*|\textbf{x},\textbf{x}_*,\textbf{y} \sim \mathcal{N}(\bar{\textbf{f}}_* | cov(\textbf{f}_*))
\end{equation}
where $\bar{\textbf{f}}_*$ and $cov(\textbf{f}_*)$ are given by \cite{Rasmussen2006}
\begin{align}
    \bar{\textbf{f}}_* &=  k(\textbf{x}_*, \textbf{x}) \left[ k(\textbf{x}, \textbf{x})+ \sigma_n^2 \textbf{I} \right]^{-1} \textbf{y} \\
    cov(\textbf{f}_*)  &= k(\textbf{x}_*, \textbf{x}_*) - k(\textbf{x}_*, \textbf{x}) \left[ k(\textbf{x}, \textbf{x})+ \sigma_n^2 \textbf{I} \right]^{-1} k(\textbf{x}, \textbf{x}_*)
\end{align}
and  $\textbf{I}$ is the $N$-by-$N$ identity matrix.

In this work, we chose the squared exponential kernel, which is defined as 
\begin{equation}
    k(\textbf{x}_i, \textbf{x}_j) = \sigma_f^2 \exp\left(-\frac{1}{2} (\textbf{x}_i-\textbf{x}_j)^{\textrm{T}} M (\textbf{x}_i-\textbf{x}_j) \right) \label{SE}
\end{equation}
with hyperparameters $\boldsymbol\theta = \{M, \sigma_f^2\}$, where $M=\textrm{diag}(\boldsymbol\lambda)^{-2}$, and $ \boldsymbol\lambda = [\lambda_1, \dots, \lambda_n]$. The hyperparameters of the covariance function in equation \eqref{SE} have a clear interpretation, which is of great importance when trying to understand the data. Specifically, the $\lambda_i$ hyperparameters play the role of characteristic length-scales indicating the degree to which a change in the $i$-th design parameter is necessary in order to realise a significant change in the output. Such a kernel thus implements the automatic relevance determination (ARD) \cite{Neal_ARD} since the inverse of the length-scale determines the relevance of an input.

\section{Data set}
Our data set consists of $15000$ high-fidelity data points that are part of the \emph{Brunel Two-Phase Flow database}. They relate the pipe length ($L$), its diameter ($d$), the mass flow rate ($G$), the inlet pressure ($P_{in}$), and the supplied heat flux ($q$) with the local two-phase HTC ($\alpha_L$). The pipes used are made of stainless-steel with circular cross sections. Two working fluids,i.e., R134a and R245fa, are considered. Data with R134a are obtained in \cite{mahmoud2013heat} and \cite{shiferaw2009flow}, details regarding the test rig can be found in \cite{huo2007r134a}, while experiments with refrigerant R245fe are presented in \cite{pike2014flow} and \cite{thesis_AlGaheeshi}.

The experimental facilities consist of the closed-loop refrigerant main circuit, data acquisition and control, and cooling and heating systems. Direct electric heating was applied to the test section: fifteen K-type thermocouples were soldered to the outside of the tube at equal distances to provide the wall temperatures, and pressure transducers were used to measure inlet and outlet temperatures and pressures. The local heat transfer coefficient at each thermocouple point was estimated by
\begin{equation}
    \alpha_L = \frac{q}{T_w - T_s}
\end{equation}
where $T_w$ is the local inner wall temperature, $T_s$ is the local saturation temperature, and $q$ is the inner wall heat flux to the fluid. The temperature $T_s$ was deduced from the fluid pressure which, in turn, was determined based on the assumption of a linear pressure drop through the test section at the start of saturated boiling; $T_w$ was calculated by solving the one-dimensional heat conduction equation with volumetric heat generation in cylindrical coordinates. This makes $T_w$ a function of the outside surface temperature recorded by the thermocouples, the heat flux, and the tube wall thermal resistance. Further details on the calculation procedure are available in \cite{mahmoud2011surface}.

Before training, the data points are pre-processed by averaging the value of the local two-phase HTC along the pipe length. This yields approximately $1000$ points associating the input design parameters ($L$, $d$, $G$, $P_{in}$, $q$) with the average HTC, denoted by $\alpha$. Finally, we note that all the training performed in this study takes only a few seconds on a standard laptop, therefore,  the computational cost is negligible. Unlike more traditional ML tasks, where scalability with the size of the data set is a primary concern \cite{goodfellow2016deep}, in engineering applications the main challenge often consist in making the most out of a relatively small number of reliable data points.

\section{Results}
This study has two main thrusts. The first is to characterize the prediction capabilities of the model using data for both refrigerants, and study its performance as a function of the ratio between training and test points. The second is to characterize the model's performance when extrapolating. This means that the test points are chosen at the edges of the input space, as opposed to being randomly sampled.  
The performance of the models are evaluated by the mean absolute error
\begin{equation}
    MAE = \frac{1}{N} \sum_{i=1}^{N} \frac{| \alpha_{p,i} - \alpha_{e,i} |}{\alpha_{e,i}} \label{MAE}
\end{equation}
where $\alpha_{e,i}$ is the average HTC obtained from the $i$-th data point. The corresponding mean prediction by the GPR model is denoted by $\alpha_{p,i}$. The specifications \emph{tot} and \emph{test} after the $MAE$ in figures \ref{fig:1} and \ref{fig:2} refers to whether equation \eqref{MAE} was evaluated on the full data set or on test points only, respectively.

\subsection{Data efficiency of GPRs models} 
To study the performance of the model as a function of the ratio between the size of the training and test sets, the data points related to the two refrigerants are merged. No labels related to the working fluid are provided, i.e., the GPR model is unaware of which data point is obtained with which fluid, and the test points are selected randomly. Figure \ref{fig:1} shows the parity plots relating the value of $\alpha$ measured experimentally with the value predicted by the mean of the GPR model. The plots shown in figure \ref{fig:1}(a)-(c) refer to three different models trained by setting the percentage of data points used for testing to $70\%$, $50\%$, and $30 \%$, respectively.

As shown in figure \ref{fig:1}(a), the use of a test set of $70\%$ randomly sampled data points, corresponding to just $30\%$ of the data used for training, already provides a reasonable prediction, corresponding to $MAE=23.1\%$ on the test data and $16.6\%$ on the full data set. The performance significantly increases with the size of the training set: for the common choice of $30\%$ of the data used for testing, the $MAE$ on the test set is reduced to $11.5\%$, and to $3.4\%$ on the full data set, with nearly all the predictions within the dashed $\pm 30 \%$ error bounds (see figure \ref{fig:1}(c). The results shown in figure \ref{fig:1}, in particular those with the $70\%$ test set, demonstrate that our GPR models are  data efficient. 

\subsection{Automatic trade-off between data-fit and model complexity} 
\begin{figure}[!ht]
	\begin{center}
		\includegraphics[scale= 0.27]{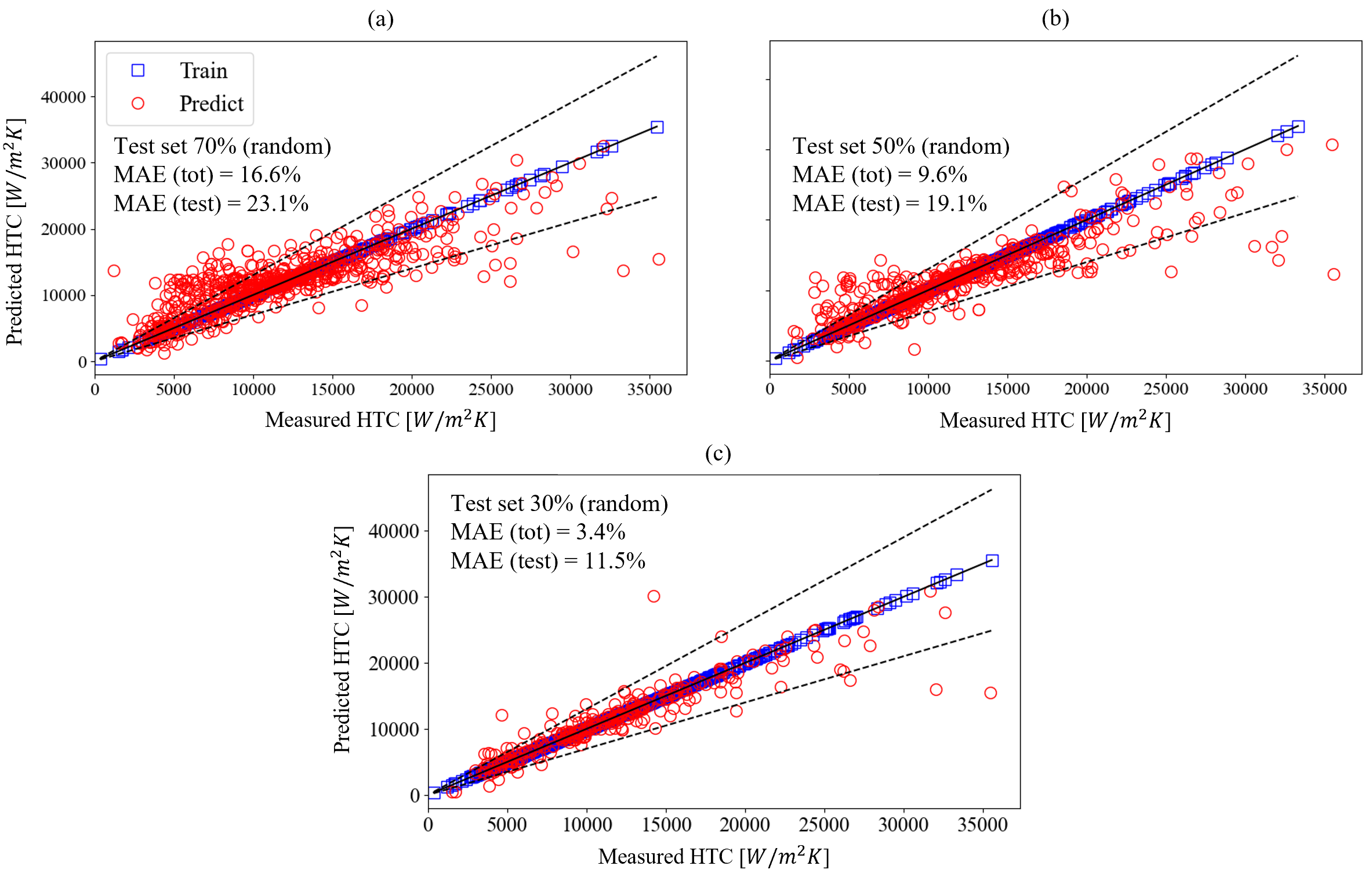}
		\caption{Randomly selected test points chosen from the datasets associated with refrigerants R134a and R245fa: parity plots showing a comparison of the experimental results and the model predictions for the cases with $70\%$ (a), $50\%$ (b), and $30\%$ (c) of test points.} \label{fig:1}
	\end{center}
\end{figure}

The parity plot in figure \ref{fig:1}(a) shows an excellent data fit of the training set, as all the blue squares lie on the parity line. 
The good fitting on the test set (red circles) of the model trained with only $30\%$ of the data gives a strong indication that the model does not overfit the data.

We remark that, unlike parametric models such as deep artificial neural networks, GPR models provide an intrinsic trade-off between data fit and model complexity \cite{Rasmussen2006}. In other words, the optimization problem in equation \eqref{training} is such that the optimal solution favours the simpler "explanation" of the data that offers the best fit. This avoids the problem of having to adjust the many hyperparameters of the ML algorithm to avoid overfitting (e.g., number and width of layers).   


\subsection{Extrapolation performance} 
\begin{figure}[!ht]
	\begin{center}
		\includegraphics[scale= 0.19]{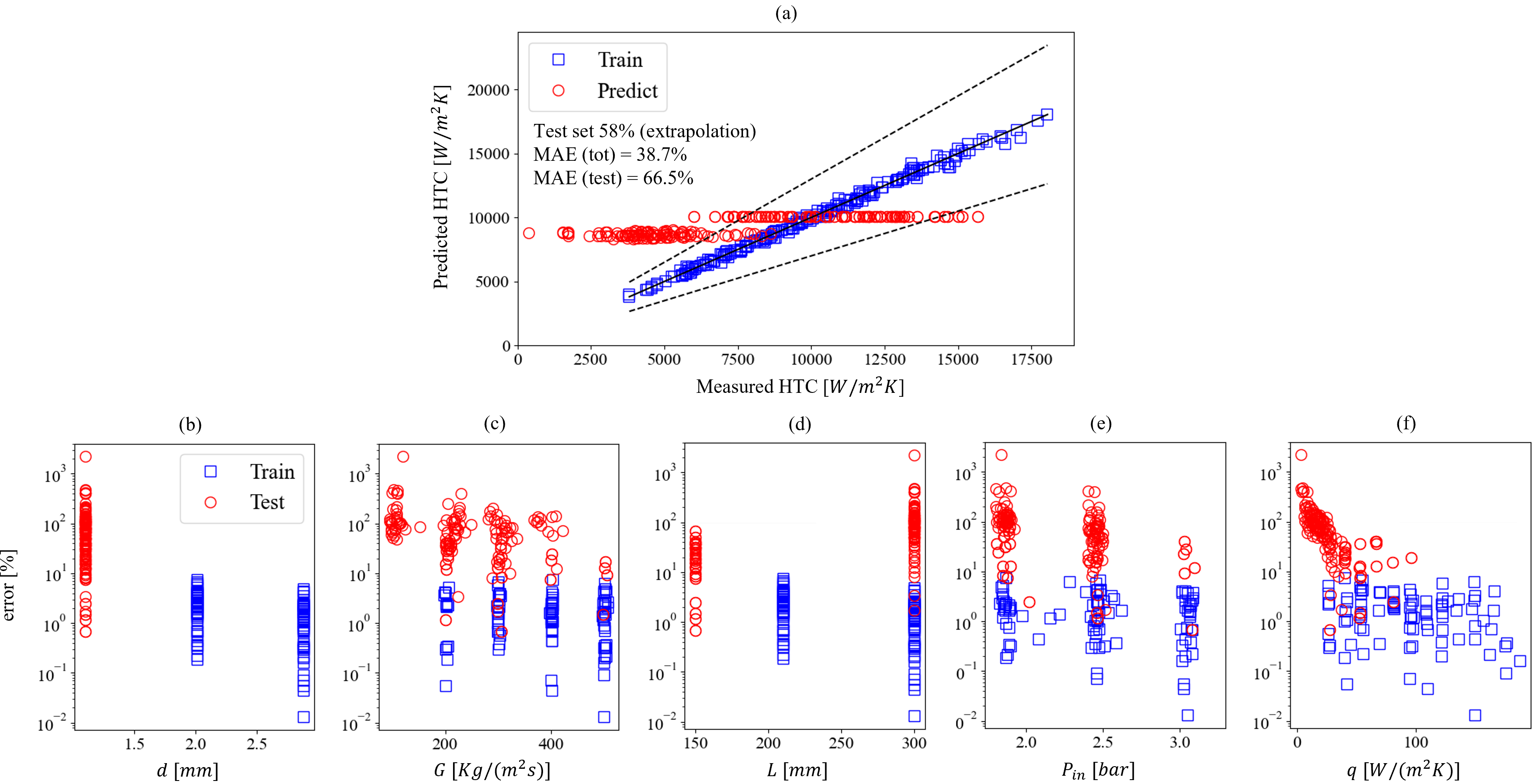}
		\caption{Using the GPR model for extrapolation using test points associated with the refrigerant R245fa: (a) parity plot showing a comparison between the experimental results and the model predictions based on 58\% of the dataset used for testing; (b)-(f) percentage error (defined in the main text) of the model prediction relative to the measured value as a function of the five design parameters considered, $d$, $G$, $L$, $P_{in}$, and $q$, respectively.} \label{fig:2}
	\end{center}
\end{figure}

We further test the extrapolation abilities of the model by deliberately selecting test points at the edges of the experimental design space, that is, we ask the model to extrapolate on all the test points. To avoid extrapolating also on different refrigerants, we restrict the training set to the data points relative to R245fa. Then, we exclude from the training set all the data points available for the smaller pipe diameter, $d=1.1$ mm, resulting in $58\%$ points used for testing. Figure \ref{fig:2}(a) shows the parity plot associated with this case, while figure \ref{fig:2}(b)-(f) shows the percentage error of the model prediction relative to the measured value, $err=100|\alpha_{p,i} - \alpha_{e,i}|/\alpha_{e,i}$, as a function of the design parameters. As shown in figure \ref{fig:2}(c)-(f), by eliminating smaller diameters we  extrapolate also along other design parameters, i.e., the mass flow rate ($G$),  pipe length ($L$), inlet pressure ($P_{in}$), and the heat flux ($q$). 

As figure \ref{fig:2} (a) shows, the $MAE$ is nearly $66.5\%$ on the test set, and $38.7\%$ on the full data set. This result is  unsurprising considering the large amount of test data compared to the total, and that the model is extrapolating along four out of five input directions. Despite this, the model still favours the simplest answer, namely it predicts nearly the same value of $\alpha$ for all the test points, providing a prediction within the dashed $\pm 30 \%$ error lines for approximately half of the test points. As discussed in the next section, we expect to improve the extrapolation performance by incorporating part of our physical knowledge into the model prior mean.  

\section{Conclusions and future work}
%
In this work, we demonstrate the potential of Gaussian Process Regression (GPR) models in the prediction of the two-phase heat transfer coefficients in micro-pipes. We  start by investigating how the performance of GPR models changes with the size of the training set. Our models are found to perform well even when small fractions of the data set are used for training. Being data-efficient makes this approach particularly suitable for engineering applications due to the large cost of obtaining high-fidelity data. Regardless of the data used for training, the model does not overfit. This automatic trade-off between the model's capacity (i.e., data-fit) and complexity is a key feature to facilitate the deployment of GPR models. Practically, this allows the end user to employ the same algorithm to train different models on different data set, thus simplifying the job of carefully hand-tuning the model's hyperparameters when either the task or the data change, a process for which clear and general guidelines are case-dependent.

Finally, we show how the model's performance significantly deteriorates when it is tested on data at the edges of the design space, i.e., when we try to extrapolate, an issue not unique to GPR models. Still, we argue that the framework of GPR has the potential to handle the task of extrapolating more successfully. While we leave this for future work, we observe how a first improvement could be obtained by using existing correlations for the HTC as prior mean of the GPR. In the current implementation the prior mean of the GPR is assumed to be zero, meaning that we are not exploiting our prior knowledge about the system. Other current developments include the use of multi-output (or vector-valued) GPR models to extrapolate to different fluids or other categorical (i.e., not continuous) design variables. 

%
%
\section*{Acknowledgments}
We acknowledge funding from the Engineering and Physical Sciences Research Council, UK, through the grant Boiling Flows in Small and Microchannels (BONSAI): From Fundamentals to Design (EP/T03338X/1, EP/T033045/1), the
Programme Grant PREMIERE (EP/T000414/1), and the Alan Turing Institute. L. Magri acknowledges financial support from the ERC Starting Grant PhyCo 949388.

%

\section*{Nomenclature}.
\vspace{-0.2in}
\begin{table}[htbp]
	\begin{tabular}[t]{p{1cm} p{4cm} l}
        $\alpha_L$ & {local HTC} & [${\rm W}/({\rm m}^2 {\rm K})$]\\
		$\alpha$ & {HTC averaged along pipe} & [${\rm W}/({\rm m}^2 {\rm K})$]\\
		$\alpha_p$ & {predicted HTC}	& [${\rm W}/({\rm m}^2 {\rm K})$]\\
        $\alpha_e$ & {experimental HTC}	& [${\rm W}/({\rm m}^2 {\rm K})$]\\
        $d$ & {pipe diameter} & [mm] \\
        $\epsilon$ & {output noise} & [${\rm W}/({\rm m}^2 {\rm K})$] \\
        $err$ & {relative prediction error} & [-] \\
        $f$ & {surrogate function} & [${\rm W}/({\rm m}^2 {\rm K})$] \\
        $\textbf{f}_*$ & {GPR posterior prediction} & [${\rm W}/({\rm m}^2 {\rm K})$] \\
        $\bar{\textbf{f}}_*$ & {GPR mean prediction} & [${\rm W}/({\rm m}^2 {\rm K})$] \\
        $G$ & {mass flux} & [${\rm kg}/({\rm m}^2 {\rm s})$] \\
        $GP$ & {Gaussian process} &  \\
        ${\rm \textbf{I}}$ & {N-by-N identity matrix} &  \\
        $k$ & {covariance function} & \\
        $\boldsymbol{\lambda}$ & {characteristic length-scales} &  \\
        $L$ & {pipe length} & [mm] \\
        $m$ & {mean of GPR} & [${\rm W}/({\rm m}^2 {\rm K})$]\\
        $MAE$ & {mean absolute error} & \\
	\end{tabular}
\hfill
	\begin{tabular}[t]{p{1cm} p{4cm} l}
         $n$ & {dimensions of GPR's input space} & \\
        $N$ & {number of observations} &  \\
            $\mathcal{N}$ & {Gaussian distribution} &\\
            $P_{in}$ & {inlet pressure} & [bar] \\
		$q$ & {heat flux} & [${\rm W}/({\rm m}^2 {\rm K})$] \\
            $\mathbb{R}$ & {real numbers} &  \\
            $\sigma_f$ & {kernel's variance} &  \\
            $\sigma_n$ & {noise's variance} & [${\rm W}/({\rm m}^2 {\rm K})$]  \\
	    $\theta$ & {kernel's hyperparameters} & \\
            $\hat\theta$ & {GPR's hyperparameters} & \\
            $T_s$ & {saturation temperature} & $ [{\rm K}$] \\
            $T_w$ & {wall temperature} & [${\rm K}$] \\
            $\mathcal{X}$ & {GPR's input space} & \\
            $\textbf{x}_i$ & {$i$-th input location} & \\
            $\textbf{x}_*$ & {test input location} & \\
            $\textbf{X}$ & {All input locations} & \\
            $\textbf{y}_i$ & {$i$-th output} & [${\rm W}/({\rm m}^2 {\rm K})$] \\
	\end{tabular}
\end{table}


\bibliography{My_Collection_1}

\end{document}